\documentclass{cs19proc}

\usepackage{url}

\editors{G.~A. Feiden}
\publisher{Zenodo}
\conference{The 19th Cambridge Workshop on Cool Stars, Stellar Systems, and the Sun}
\conferencedate{2016}

\title{Short timescale variables in stellar clusters: From Gaia to ground-based telescopes}
\author{Maroussia Roelens,$^{1}$ 
        Sergi Blanco-Cuaresma,$^{1}$
        Laurent Eyer,$^{1}$
        Nami Mowlavi,$^{1}$
        Isabelle Lecoeur-Ta\"{i}bi,$^{2}$
        Lorenzo Rimoldini,$^{2}$
        Lovro Palaversa,$^{1}$
        Maria S\"{u}veges,$^{2}$
        Jonathan Charnas$^{2}$}

\affiliation{$^{1}$ Observatoire de Gen\`{e}ve, D\'{e}partement d'Astronomie, Universit\'{e} de Gen\`{e}ve, Chemin des Mailletes 51, CH-1290 Versoix, Switzerland \\
$^{2}$ Observatoire de Gen\`{e}ve, D\'{e}partement d'Astronomie, Universit\'{e} de Gen\`{e}ve, Chemin d'Ecogia 16, CH-1290 Versoix, Switzerland}

\shorttitle{Short timescale variables in stellar clusters}
\shortauthors{Maroussia Roelens \& Sergi Blanco-Cuaresma}

\abs{Combined studies of variable stars and stellar clusters open great horizons, and they allow us to improve our understanding of stellar cluster formation and stellar evolution. In that prospect, the Gaia mission will provide astrometric, photometric, and spectroscopic data for about one billion stars of the Milky Way. This will represent a major census of stellar clusters, and it will drastically increase the number of known variable stars. In particular, the peculiar Gaia scanning law offers the opportunity to investigate the rather unexplored domain of short timescale variability (from tens of seconds to a dozen of hours), bringing invaluable clues to the fields of stellar physics and stellar aggregates.

We assessed the Gaia capabilities in terms of short timescale variability detection, using extensive light-curve simulations for various variable object types. We showed that Gaia can detect periodic variability phenomena with amplitude variations larger than a few millimagnitudes.

Additionally, we plan to perform subsequent follow-up of variables stars detected in clusters by Gaia to better characterize them. Hence, we develop a pipeline for the analysis of high cadence photometry from ground-based telescopes such as the 1.2m Euler telescope (La Silla, Chile) and the 1.2m Mercator telescope (La Palma, Canary Islands).
\medbreak
Contact: maroussia.roelens@unige.ch, sergi.blanco@unige.ch}

\begin{document}

\maketitle

\section{Introduction}
\label{intro}

Stellar clusters are ideal environments to investigate variability. Knowing the cluster characteristics (age, distance, reddening and chemical abundancies), it is possible to derive some basic properties of its individual members exhibiting variability, as well as their evolutionary status. Such combined studies provide additional information about stellar evolution processes, with respect to analyzing variable stars alone. In the domain of asteroseismology, for example, investigating pulsating variable stars in stellar clusters allows us to better constrain the modeling of the oscillation mode, and helps refine our understanding of the instability strip.
Conversely, variable stars can contribute to the characterization of their host cluster. Hence, if a stellar cluster is confirmed to contain RR Lyrae stars or Cepheids (whose period - luminosity relation makes them ideal standard candles), then it is possible to determine the distance and reddening of the cluster.

In this prospect, the Gaia ESA cornerstone mission, launched in December 2013, offers a unique opportunity to significantly change the landscape in both domains of stellar variability and stellar aggregates. During its 5-year lifetime, Gaia will survey about one billion objects over the entire sky, down to $G \approx 20$ mag ($G$ being the Gaia broad-band white light magnitude). Gaia's science data comprise micro-arcsecond global astrometry, photometry with standard errors at the milli-magnitude (mmag) level, and medium-resolution (resolving power $\sim  11,500$) spectroscopy down to $G \approx 17$ mag \citep{deBruijne2012}\footnote{For more information on Gaia performances, please see the Gaia webpage \url{http://www.cosmos.esa.int/web/gaia/science-performance}}. This will represent a major census of stellars clusters, determining accurate distances for a few hundred of them in our Galaxy, and providing a complete list of their members.
The derived isochrones will enable to improve stellar models, and to learn more about stellar evolution, structure and atmospheres. In addition, by combining high-precision photometry with simultaneous astrometry and spectroscopy, Gaia will make a complete and homogeneous variability search possible, including low-amplitude variables. Hence, this mission will drastically increase the number of known variable stars.

The Gaia Nominal Scanning Law (NSL) involves fast observing cadences, with groups of nine consecutive CCD observations separated by about $4.85$s from each other, followed by gaps of 1h46min or 4h14min between two successive groups, where a group is referred to as field-of-view transit \citep{deBruijne2012}\footnote{For more information on the Gaia spacecraft, instruments and observing strategy, please see the Gaia webpage \url{http://www.cosmos.esa.int/web/gaia/spacecraft-instruments}}. This time sampling enables the investigation of the rather unexplored domain of \textit{short timescale variability}, from tens of seconds to a dozen hours. A variety of astronomical objects are known to exhibit such rapid variations in their light-curve, whether periodic or transient, with amplitudes ranking from a few mmags to a few magnitudes. Some examples of short timescale variable types are listed in Table \ref{tab:varTypeListPeriodic}. However, only a small number of such objects are currently known, from a few tens to a few hundred sources depending on the variable type. This is the consequence of the inherent observational constraints when targeting fast variability, both in terms of time sampling and photometric precision, particularly if you are interested in low amplitudes variations. Thanks to Gaia capabilities, these numbers are going to be significantly increased, which will improve our knowledge and understanding of these specific sources, bringing invaluable clues in several fields of astrophysics (pulsation theories, distance estimates, physics of degenerate matter...).
As part of the Gaia Data Processing and Analysis Consortium \citep[DPAC,][]{Mignard2008}, within the Coordination Unit 7 \citep[CU7,][]{Eyer2015} whose activities are dedicated to the Variability Processing, our task is to develop, implement, test and apply methods for automated detection and characterization of short timescale variables with Gaia data.

Once these Gaia short timescale variable candidates are identified, supplementary ground-based observations will be performed in order to confirm the suspected variability. A specific workpackage of CU7 is dedicated to this aspect, and their activities contribute to the validation of the Gaia data prior to any public data release. Then, after Gaia data are made available to the whole community, ground-based follow-up of the detected short timescale variables will be performed to further characterize these objects. In this double perspective, we plan to perform a ground-based high cadence photometric follow-up of short timescale variable candidates detected by Gaia and belonging to stellar clusters. Thus, we develop a pipeline for reduction and analysis of the resulting photometric data, which will be observed with the $1.2$m Euler Swiss telescope (La Silla, Chile) and with the $1.2$m Mercator Flemish telescope (La Palma, Canary Islands, Spain).

In this work, we present the preparation of this combined analysis of short timescale variability, from Gaia data to ground-based observations. In section \ref{shortTimescaleGaia}, we investigate the power of Gaia for detecting short timescale variables, as well as the question of contamination by false positives. Section \ref{ePipe} is dedicated to the description of \textit{ePipe}, our reduction pipeline for the analysis of ground-based photometric observations. We show an example of what can be done with \textit{ePipe}, emphasizing on the complementarity between data from telescopes on Earth and Gaia data.

\section{Detection of short timescale variability with Gaia}
\label{shortTimescaleGaia}

Our goal is to assess Gaia capabilities in terms of short timescale variability detection, using Gaia per-CCD photometry in $G$ band. Our study is based on extensive light-curve simulations, for various types of short timescale variables. In this proceeding, we limit ourselves to the case of fast \textit{periodic} variability, with periods shorter than $0.5$d. Further analysis including \textit{transient} variability will be detailed in a future paper (Roelens et al., in prep.).

\subsection{The variogram method}
\label{shortTimescaleGaia_variogram}

\begin{figure*}[ht]
\centering
\includegraphics[width=0.49\linewidth, page=2, trim={5cm 0cm 5cm 0cm}, clip=true]{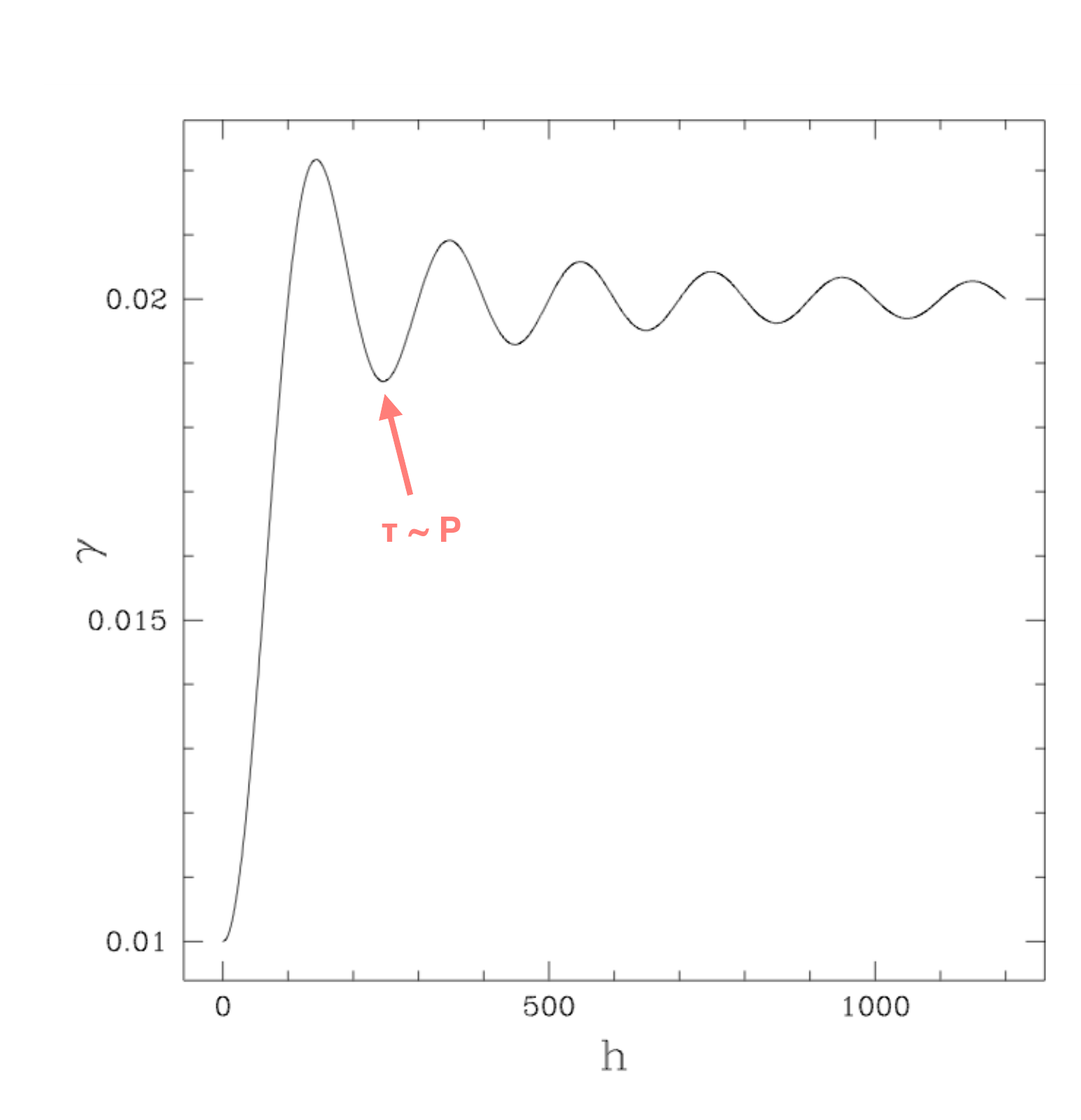}
\includegraphics[width=0.4\linewidth, page=3, trim={8cm 0cm 8cm 0cm}, clip=true]{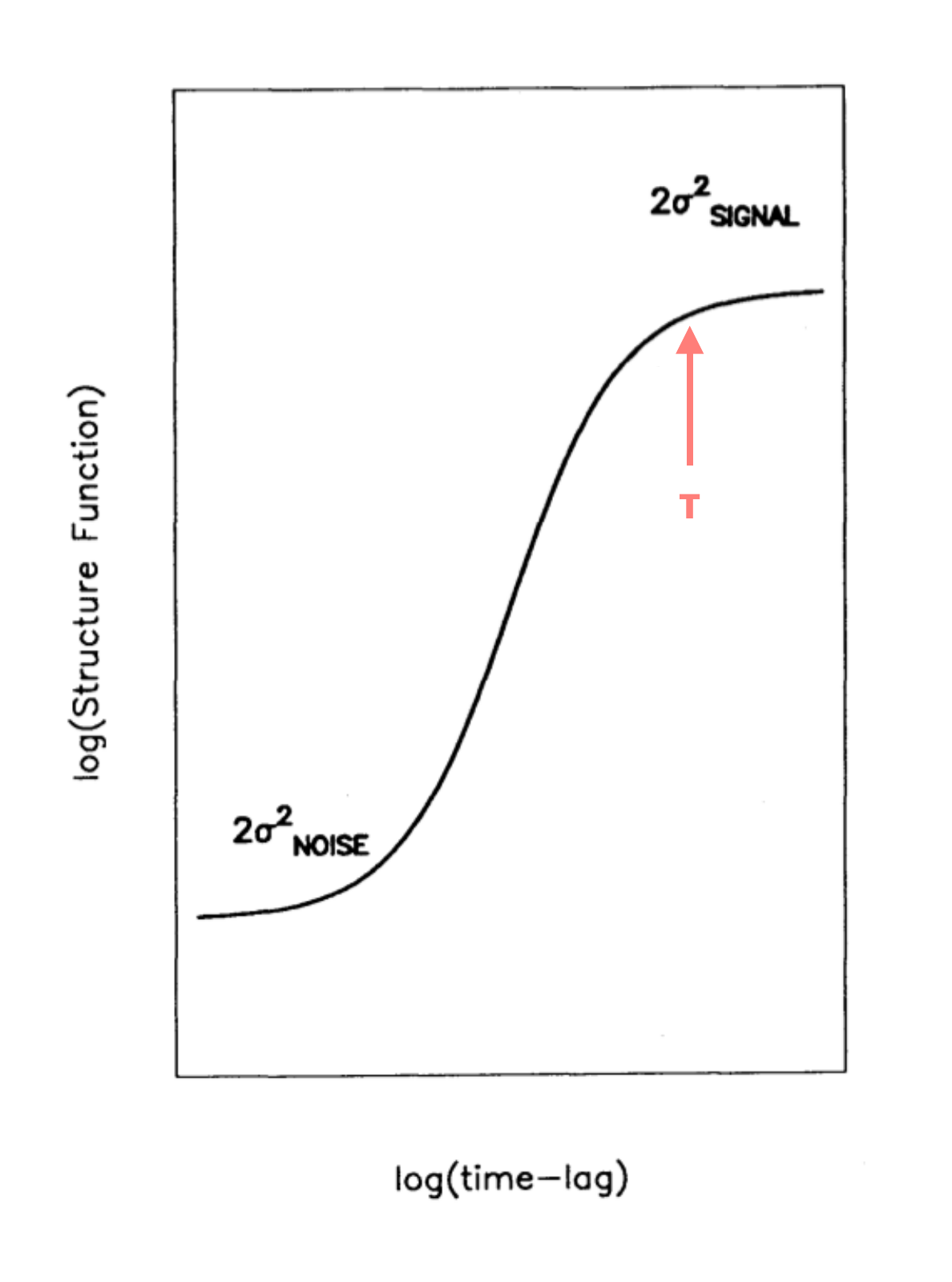}
\caption{Typical variogram plots. Left: for a periodic/pseudo-periodic variable \citep{Eyer1999}. Right: for a transient variable, only exploring lags up to the first structure characteristic of variability \citep{Hughes1992}. In each case, the feature used to estimate the typical timescale is pointed by a red arrow.}
\label{fig:typicalVariograms}
\end{figure*}

As part of our activities within CU7, we are responsible for implementing an algorithm specifically dedicated to the automated detection of short timescale variables from Gaia photometry. The method we use for this purpose is the \textit{variogram method}, also known as the \textit{structure function method}. Its basic idea is to investigate astronomical light-curves for variability by quantifying the difference in magnitude between two measurements as function of the time lag between them. Given a time series of magnitudes $(m_{i})_{i=1..n}$ observed at times $(t_{i})_{i=1..n}$. The variogram value for a time lag $h$ is denoted by $\gamma(h)$. It is defined as the average of the squared difference in magnitude $(m_{j} - m_{i})^{2}$ computed on all pairs $(i, j)$ such that $\vert t_{j} - t_{i} \vert = h$. This formulation corresponds to the classical first order structure function as defined by \citet{Hughes1992}. By exploring different lag values, a \textit{variogram plot} associated to the time series can be built (referred to as variogram). This variogram provides information on how variable the considered source is, and on the characteristics of the variability when appropriate. Figure \ref{fig:typicalVariograms} shows typical variograms for a periodic or pseudo-periodic variable, and for a transient.
If the analyzed time series exhibits some variability, the expected features in the variogram are:
\begin{itemize}
\item For the shortest lags, a plateau at $\gamma \sim 2\sigma_{noise}^{2}$, where $\sigma_{noise}$ is the measurement noise.
\item Towards longer lags, an increase of the variogram values, followed by a second plateau at $\gamma \sim 2\sigma^{2}$, where $\sigma$ is the standard deviation of the time series.
\end{itemize}
If the underlying variation is periodic or pseudo-periodic, this second plateau is followed by a succession of dips. In the case of a transient, the second plateau can be followed by complex structures, like other plateaus or decreases in the variogram values, depending on the origin of the variability (stochasticity, flares, ...).

Indeed, the power of the variogram method relies on the fact that it can handle periodic variability as well as transient events. In addition, it provides an estimate of the characteristic variation timescale. Hence, for a transient, the typical timescale $\tau$ is of the order of the lag at which the variogram starts plateauing. As mentioned previously, there can be more than one plateau, indicating that the considered variability has several typical timescales. For a periodic variable, the position of the first dip after plateauing gives a rough estimate of the underlying period. We emphasize that we do not consider the variogram as a substitute to other period search methods, such as the Fourier based periodograms, which determine the period much more precisely in the case of strictly periodic variations. But it can be complementary to these periodograms, and for example help distinguish true period from aliases \citep[e.g., see ][]{Eyer1999}. Besides the variogram performs quite well with non-strictly periodic variability, where Fourier-based methods usually fail. 
However, in this work we are only interested in the outcome of the variogram method in terms of variability detection. We do not present any result on characteristic timescale estimation. This analysis will be detailed in a future paper (Roelens et al., in prep.).

Suppose that a variogram similar to one of these in Figure \ref{fig:typicalVariograms} is derived from a light-curve observed by Gaia. We denote by $h_{k}$ the explored lags and by $\gamma(h_{k})$ the associated variogram values. The first step is to know whether the considered source is a true variable or not. In the Gaia context, several hundreds of thousands of light-curves will be investigated for short timescale variability. Thus, the detection of short timescale variable candidates should be performed in an automated way.
One possible criterion to distinguish constant sources from variable ones,  according to their variograms, is to fix a detection threshold $\gamma_{det}$ such that:
\begin{itemize}
\item If for at least one lag $h_{k}$ you have $\gamma(h_{k}) > \gamma_{det}$ then your source is flagged as variable,
\item If for all lags you have $\gamma(h_{k}) < \gamma_{det}$ then your source is flagged as constant.
\end{itemize}
$\gamma_{det}$ defines the variance value above which you consider that the variability in the signal is sufficiently significant not to be due to noise only. Depending on the chosen value, the definition of what is "real" variability is more or less restrictive.
For the sources identified as variable, we define the \textit{detection timescale} $\tau_{det}$ as the smallest lag for which 
$\gamma(\tau_{det}) > \gamma_{det}$. $\tau_{det}$ is characteristic of the underlying variability in the light-curve, and quantifies the associated variation rate. The shorter the detection timescale, the higher the change in magnitude per unit time.

\subsection{Simulated short timescale variable light-curves}
\label{shortTimescaleGaia_simus}

In order to evaluate the power of the variogram method for detecting short timescale variables with Gaia, we decided to simulate extensive light-curve sets for different types of such astronomical objects. The main purpose of these simulations is to mimic how variability is seen through the eyes of Gaia. As mentioned before, we focused only on periodic variables. The short timescale variable types that we simulate are listed in Table \ref{tab:varTypeListPeriodic}, together with the corresponding period and amplitude ranges. Note that the word "amplitude" refers to the peak-to-peak amplitude of the variation. In this work, all the simulated variables have periods shorter than $0.5$d.

\begin{figure*}[ht]
\centering
\includegraphics[width=0.49\linewidth,page=22]{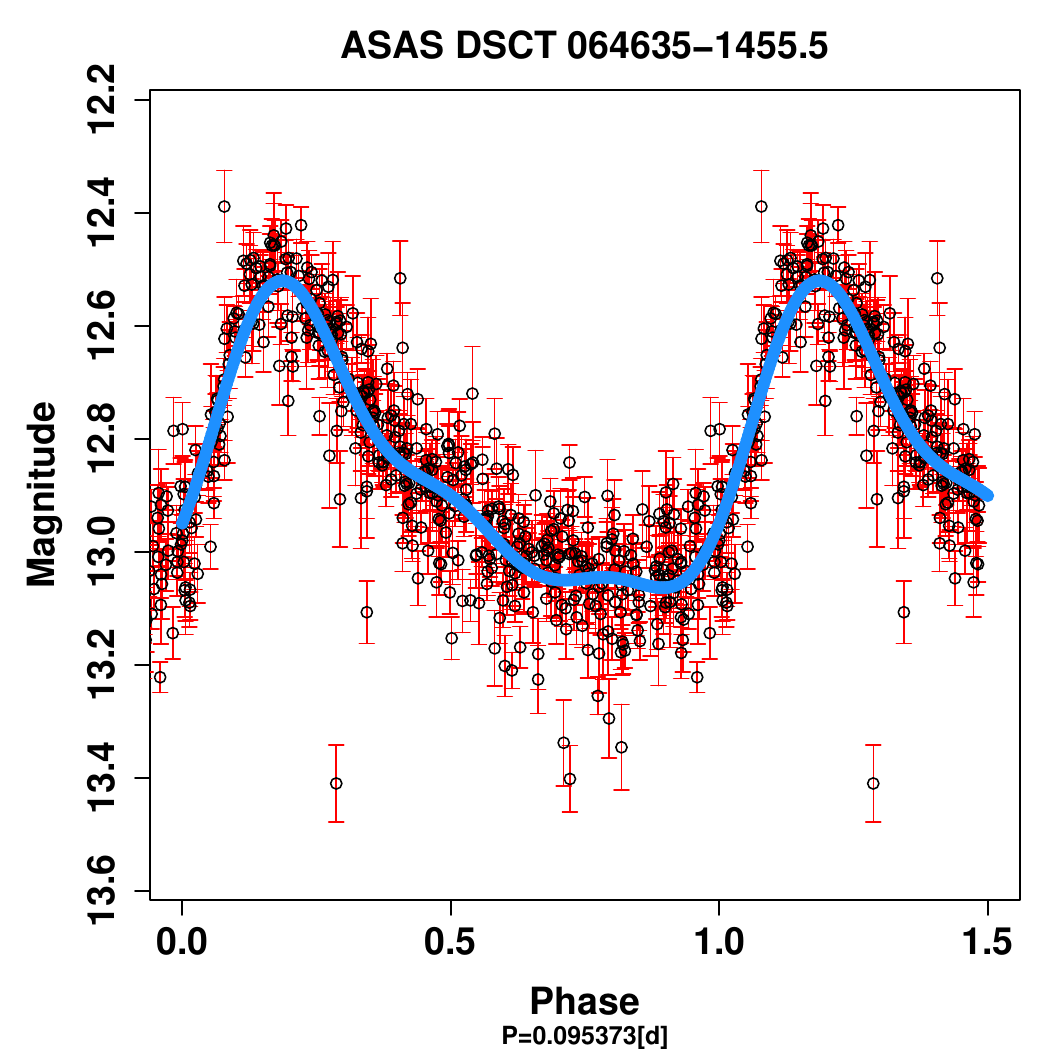}
\caption{Example of $\delta$ Scuti template, from ASAS-3 V band measurements. The black circles with red error bars correspond to the observed ASAS light-curve. The empirical template is overplotted in blue.}
\label{fig:dsctTemplate}
\end{figure*}

\begin{figure*}[ht]
\centering
\includegraphics[width=0.49\linewidth, page=6, trim={0 0 0 1.5cm}, clip=true]{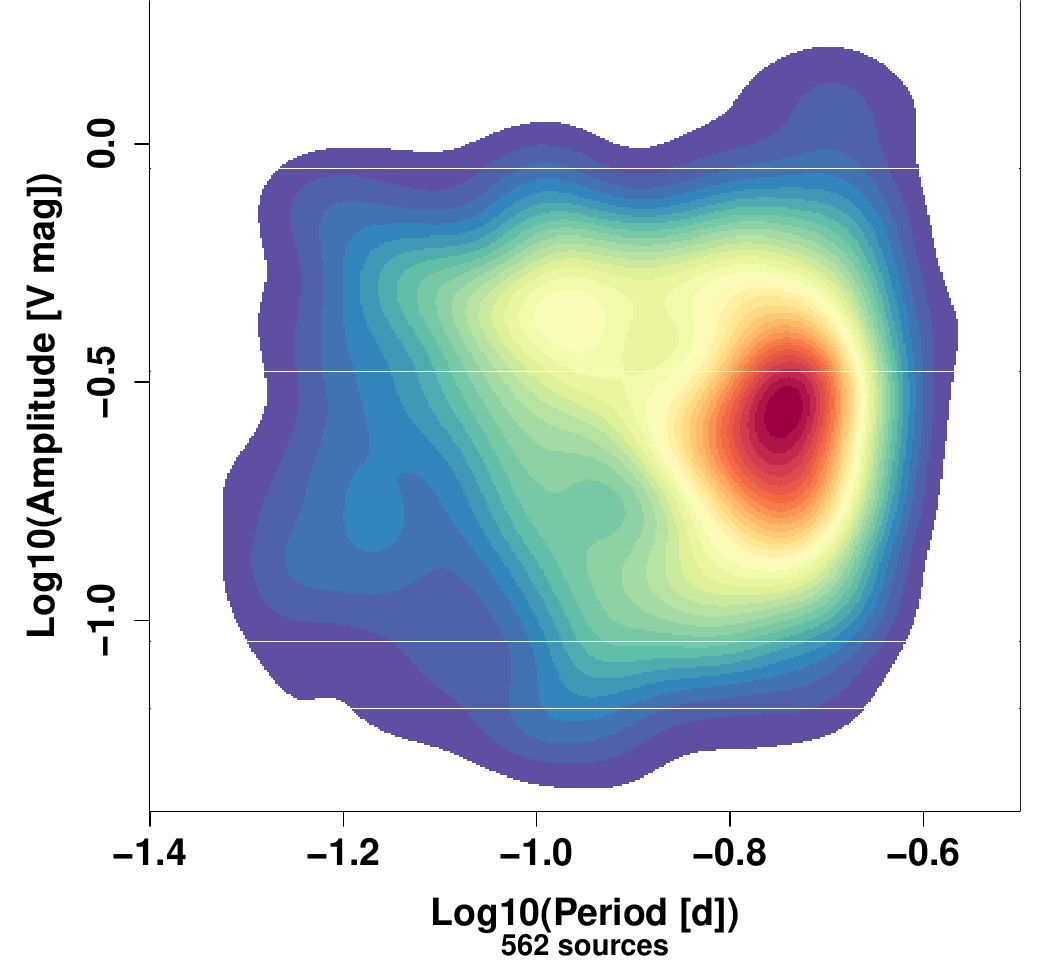}
\caption{Probability distribution in the period-amplitude diagram for $\delta$ Scuti stars, from ASAS-3 catalogue of variable stars.}
\label{fig:dsctPAdistrib}
\end{figure*}

The simulation principle we used for generating periodic light-curves is the following one. First, we build empirical, phase-folded, normalized templates, from observed light-curves found in the literature with relevant period and amplitude information. Figure \ref{fig:dsctTemplate} represents an example of a $\delta$ Scuti template, retrieved from ASAS-3 catalogue\footnote{\url{http://www.astrouw.edu.pl/asas/?page=catalogues}} of variable stars \citep{Pojmanski2002}. The second ingredient of our recipe is the magnitude of the source we want to simulate. We take it randomly between $8$ and $20$ mag, which are approximately the bright and faint end of the Gaia $G$ photometry for most objects of interest. Finally, we choose the period $P$ and amplitude $A$ of the simulated variable star. To make our simulations realistic, when possible we draw the couple ($P$, $A$) from empirical period-amplitude probability distributions, retrieved from existing variable star catalogues. An example of such a 2D probability distribution can be found in Figure \ref{fig:dsctPAdistrib}. If there is not enough information in literature, we randomly draw the period and amplitude in the appropriate ranges given in Table \ref{tab:varTypeListPeriodic}. Once we have all these ingredients, we scale the phase-folded template at the proper amplitude A. Then we generate the observing time values, following the appropriate time sampling and over the required time step, and we convert them into phases according to the simulation period P. Finally, we compute the magnitudes corresponding to each observing time from the scaled template.

\begin{figure*}[ht]
\centering
\includegraphics[width=0.49\linewidth, page=2, trim={0 0 0 1.2cm}, clip=true]{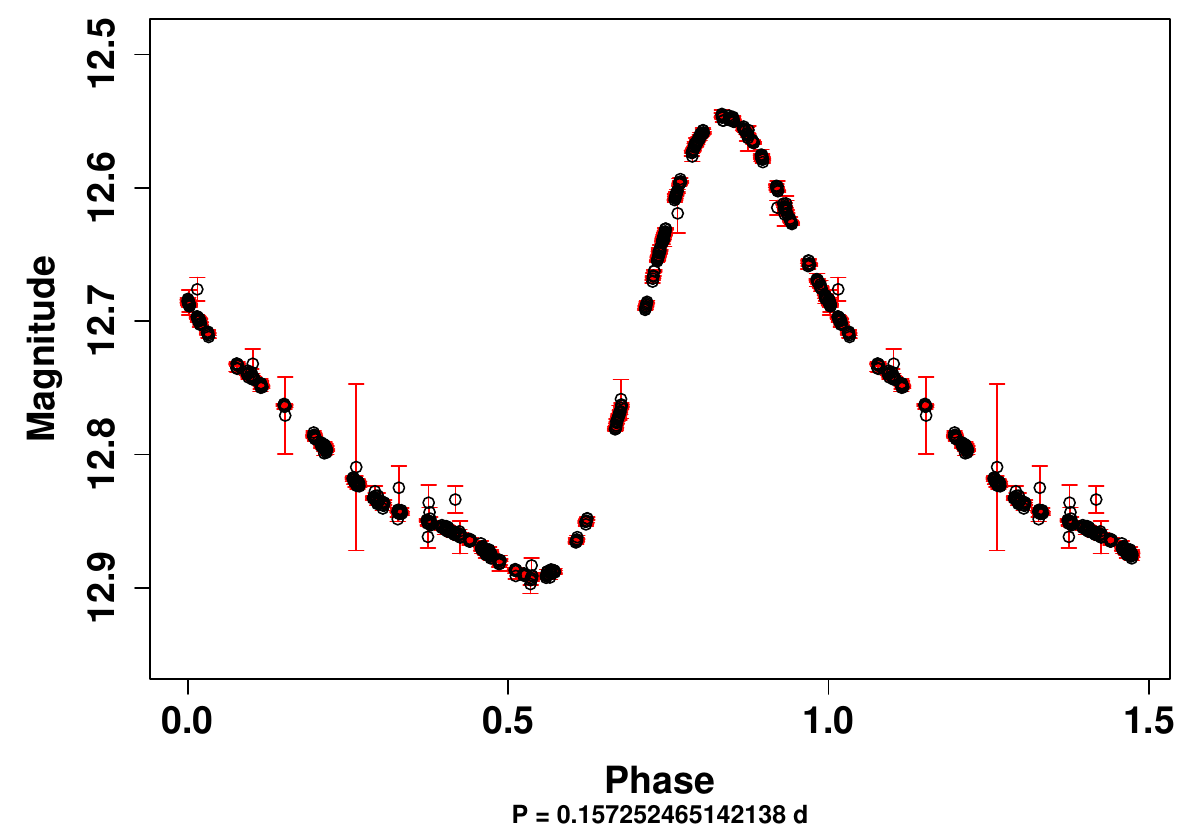}
\includegraphics[width=0.49\linewidth,page=3, trim={0 0 0 1.2cm}, clip=true]{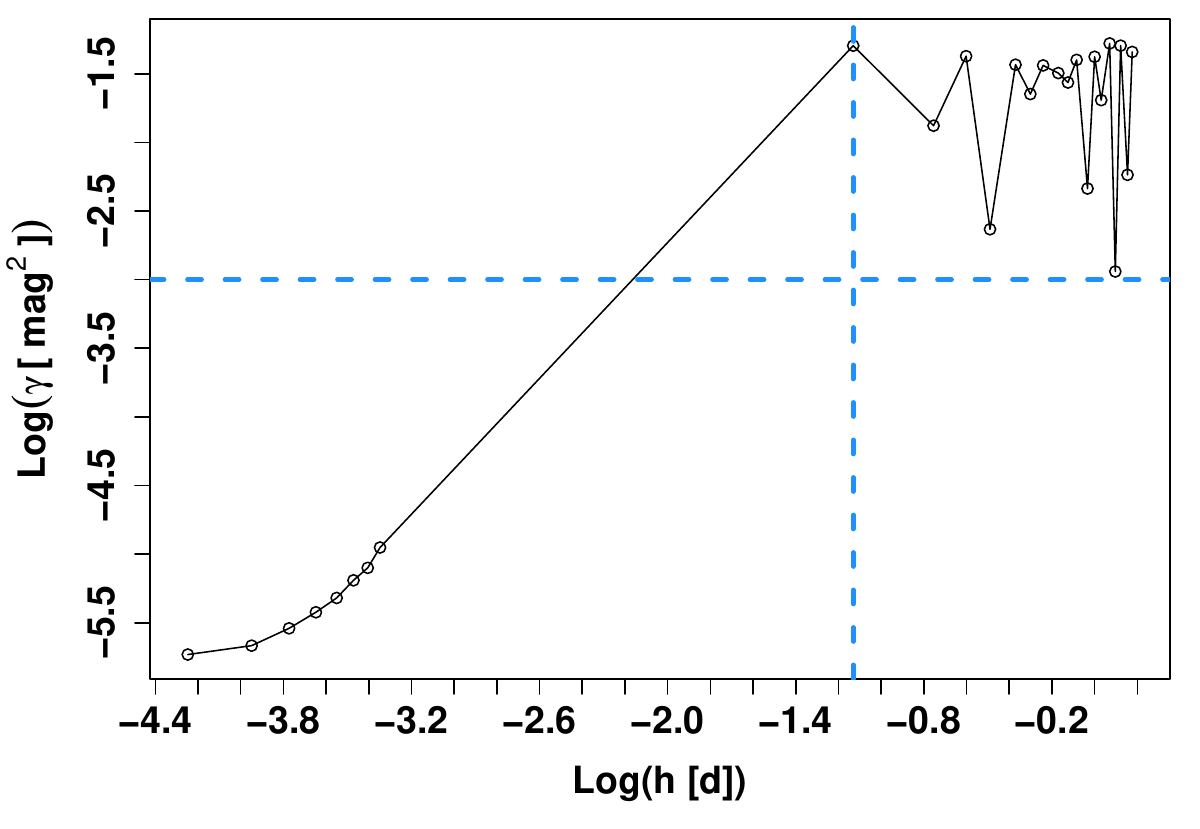}
\caption{Example of $\delta$ Scuti Gaia-like light-curve (left), and corresponding observational variogram (right). The blue dashed lines indicate the detection threshold ($\gamma_{det} = 10^{-3} \mathrm{mag}^{2}$) and the associated detection timescale $\tau_{det}$.}
\label{fig:dsct84gaialike}
\end{figure*}

For our analysis, we generate two different types of light-curves:
\begin{itemize}
\item The \textit{continuous} light-curves, noiseless, with a very tight and perfectly regular time sampling, over a timespan $\Delta t \sim 5P$ where $P$ is the period of the simulation. The continuous data set is used to assess what we could detect in an ideal situation. It comprises 100 distinct simulations for each of the 8 variable types listed in Table \ref{tab:varTypeListPeriodic}. 
\item The \textit{Gaia-like} light-curves, corresponding to the same variables as in the continuous data set (same period, amplitude and magnitude), but this time with a time sampling following the Gaia NSL, and adding noise according to a magnitude-error distribution retrieved from real Gaia data. The timespan is of the order of five years, which is the nominal duration of the Gaia mission.
\end{itemize}
The left panels of Figures \ref{fig:dsct84continuous} and \ref{fig:dsct84gaialike} show the two light-curves obtained for the same $\delta$ Scuti star, one simulated in the continuous way, and the other simulated in the Gaia-like way.

\subsection{Variogram analysis on the continuous data set}
\label{shortTimescaleGaia_analyseCont}

\begin{figure*}[ht]
\centering
\includegraphics[width=0.49\linewidth, page=2, trim={0 0 0 1.2cm}, clip=true]{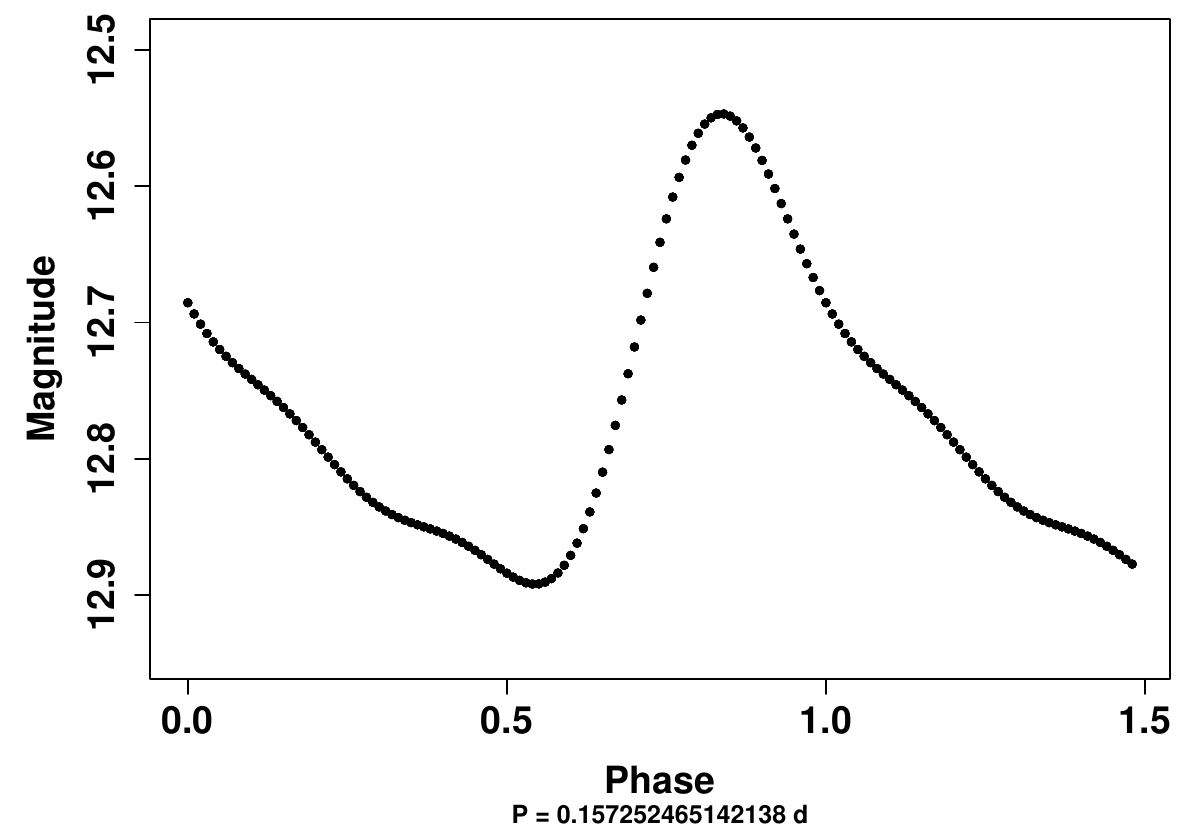}
\includegraphics[width=0.49\linewidth, page=3, trim={0 0 0 1.2cm}, clip=true]{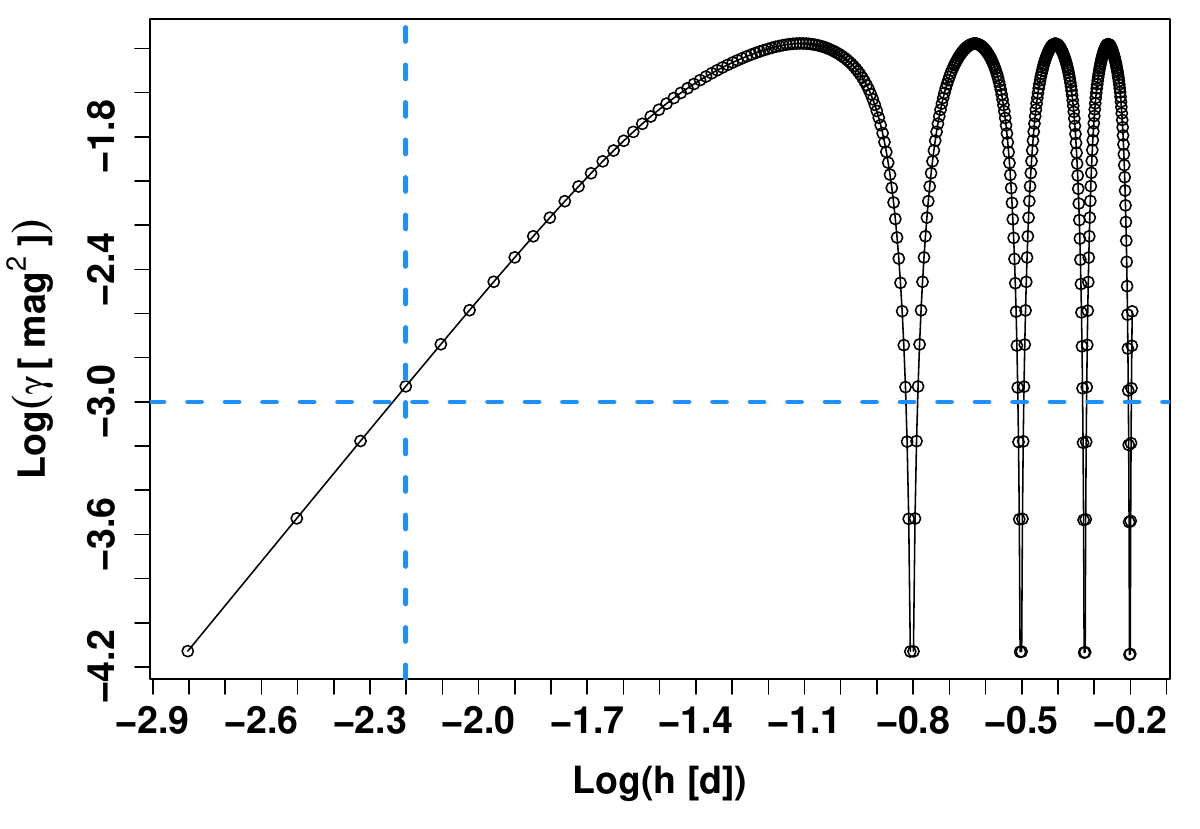}
\caption{Example of $\delta$ Scuti continuous light-curve (left), and corresponding theoretical variogram (right). The blue dashed lines indicate the detection threshold ($\gamma_{det} = 10^{-3} \mathrm{mag}^{2}$) and the associated detection timescale $\tau_{det}$.}
\label{fig:dsct84continuous}
\end{figure*}

For each simulated continuous light-curve, we calculated the associated theoretical variogram, for the appropriate lag values defined by the underlying time sampling (i.e. explored lags are multiples of the time step $\delta t$). Figure \ref{fig:dsct84continuous} represents an example of such light-curve and variogram. Then, we applied the short timescale variability detection criterion described in section \ref{shortTimescaleGaia_variogram}, with $\gamma_{det} = 10^{-3}\mathrm{mag}^{2}$. As shown in right panel of Figure \ref{fig:dsct84continuous}, our $\delta$ Scuti example is detected, with a detection timescale $\tau_{det,continuous} \simeq 10^{-2.2} \mathrm{d} \simeq 9.1\mathrm{min}$.

\begin{figure*}[ht]
\centering
\includegraphics[width=0.49\linewidth, page=43, trim={0 0 0 3cm}, clip=true]{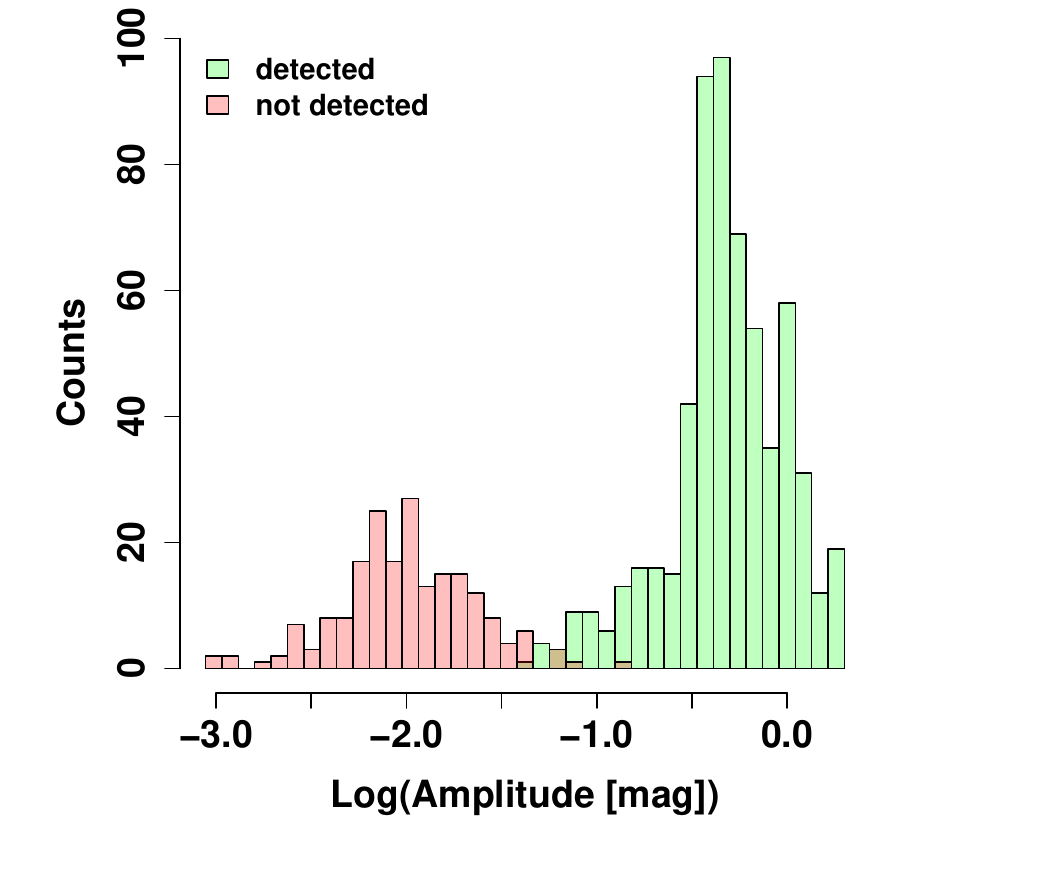}
\includegraphics[width=0.49\linewidth, page=79, trim={0 0 0 3cm}, clip=true]{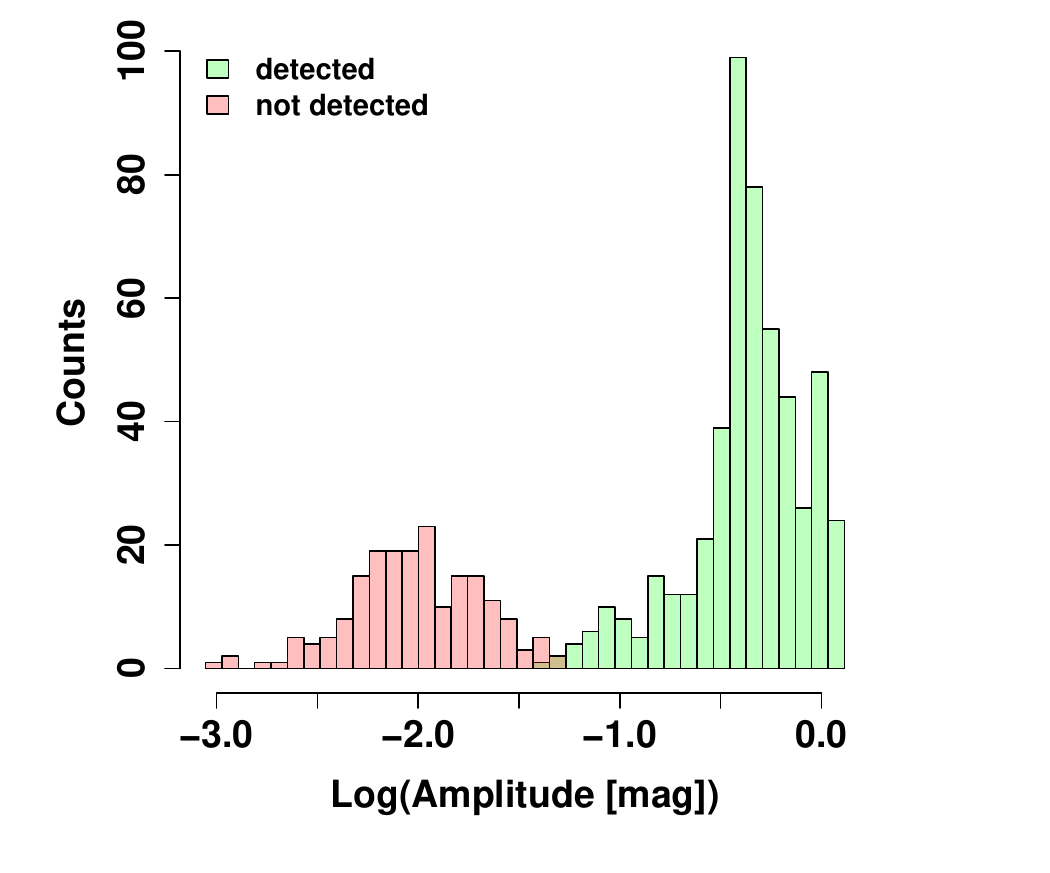}
\includegraphics[width=0.49\linewidth, page=75, trim={0 0 0 3cm}, clip=true]{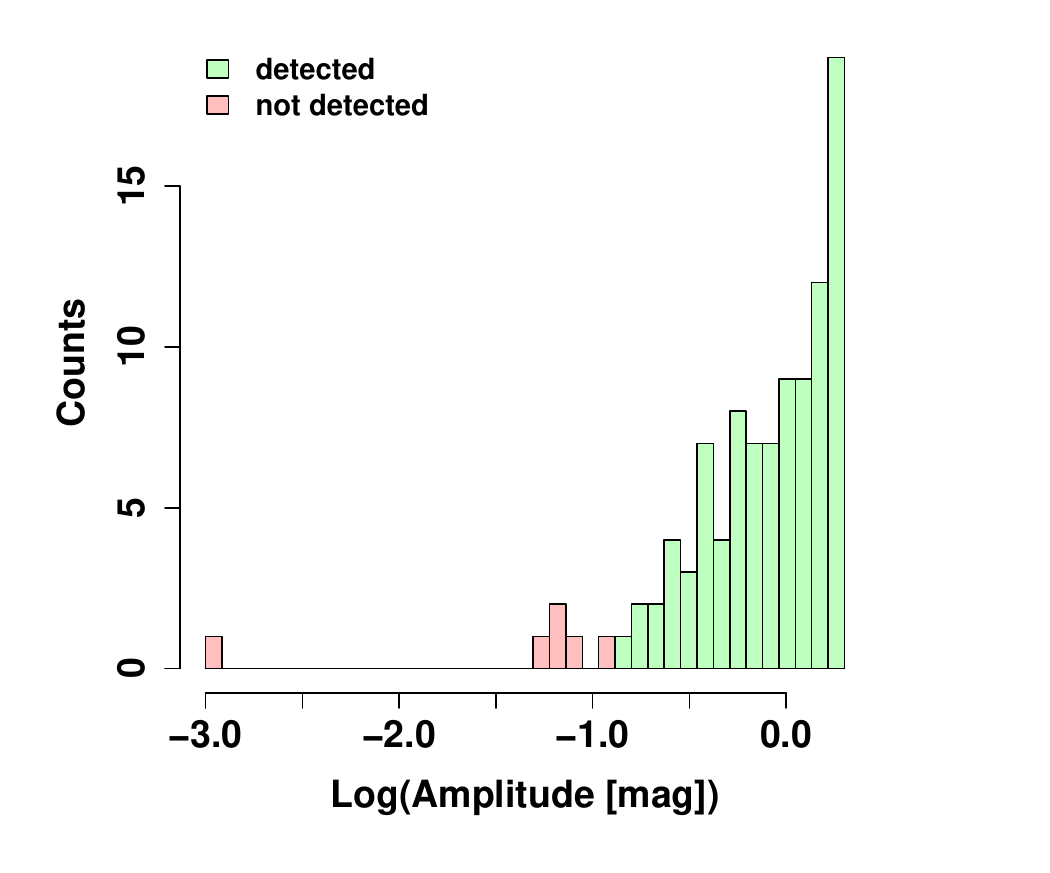}
\caption{Distribution of input amplitude A for the periodic variable simulations, separating the detected (green) and the not detected (red) sources, continuous data set. Top left: all variable types. Top right: all variable types but AMCVn. Bottom: only the AMCVn simulations.}
\label{fig:histAcontinuous}
\end{figure*}

Among the 800 periodic variable simulations in the continuous data set, 603 (75.4\%) are flagged as short timescale variables with our criterion, and 197 (24.6\%) are missed. The main discriminating criterion between the detected and the not detected objects is the input amplitude $A$ of the simulation (see Figure \ref{fig:histAcontinuous}). With our choice of $\gamma_{det}$, the smallest amplitude detected is $A \simeq 0.043\mathrm{mag}$. The overlap visible between the two distributions, in the top-left panel of Figure \ref{fig:histAcontinuous}, mostly corresponds to AM CVn simulations. Indeed, AM CVn eclipses are very fast. Thus, in our simulations we have much fewer measurements sampling the eclipses than the quiescence stage, which produces variogram values lower than expected. Consequently, higher amplitudes are required for AM CVn stars to be flagged as variable than for the other variable types. This effect is clearly visible if we compare the amplitude distributions of detected and not detected variables, for AM CVn simulations only (top-right panel of Figure \ref{fig:histAcontinuous}), and for all simulated types but AM CVn (bottom panel of Figure \ref{fig:histAcontinuous}). All in all, with $\gamma_{det} = 10^{-3} \mathrm{mag}^{2}$, the amplitude limits for detecting short timescale variables in an ideal noiseless case are:
\begin{itemize}
\item $A \gtrsim 0.14$ mag for AM CVn stars,
\item $A \gtrsim 0.046$ mag for the seven other types simulated.
\end{itemize}

\subsection{Variogram analysis on the Gaia-like data set}
\label{shortTimescaleGaia_analyseGaia}

Our analysis of the continuous data set (section \ref{shortTimescaleGaia_analyseCont}) reveals that, with the detection criterion applied, ideally we should detect any short period ZZ Ceti, $\beta$ Cephei, $\delta$ Scuti, RR Lyrae and eclipsing binary with amplitude above $0.046$ mag, and any AM CVn star with amplitude above $0.14$ mag. But what happens if we move to the Gaia-like sampling?
Similarly to what we did for the continuous data set, we computed the variogram associated with each simulated Gaia-like light-curve. This time, the lags explored are defined by the Gaia NSL, namely the time intervals between CCD measurements ($4.85$s, $9.7$s, $14.6$s, $19.4$s, $24.3$s, $29.2$s, $34$s, $38.8$s), and these between the different transits ($1$h$46$min, $4$h$14$min, $6$h, $7$h$46$ etc...), up to $h \simeq 1.5$d. Note that no lag can be explored from about 40s to 1h46min, which may have consequences on the detectability of some sources and on their detection timescale. Then, we applied our detection criterion, keeping $\gamma_{det} = 10^{-3}\mathrm{mag}^{2}$.
 Figure \ref{fig:dsct84gaialike} shows the Gaia-like light-curve and associated observational variogram, for the same simulated $\delta$ Scuti as in Figure \ref{fig:dsct84continuous}. As it can be seen, this $\delta$ Scuti star is also detected in the Gaia-like data set, with $\tau_{det,Gaia-like} \simeq 1$h$46$min. The detection timescale is longer in the Gaia-like case than in the continuous case, which is somehow expected: indeed, $\tau_{det,continous}$ falls in the lag gap mentioned above, hence in the Gaia-like framework the detection is pushed towards longer lags.

Among the 800 Gaia-like simulations of short period variables, 606 (75.8\%) are detected, and 194 (24.2\%) are missed. In Table \ref{tab:continuousVSgaialike}, we compare the detection results in the continuous data set to the detection results in the Gaia-like data set. Most of what we expected to detect from the ideal case is detected in the Gaia-like data set. Similarly, most of what should not be detected according to the continuous data set analysis is not detected in the Gaia-like data set. Only a few sources are flagged variable in one data set and not in the other. After further investigation of these specific cases, we found that the three sources detected in the continuous data set and not in the Gaia-like one are all AM CVn stars, with very few points sampling their eclipses in the Gaia-like light-curves. Hence their variability is missed by Gaia. Moreover, the six sources not detected in the continuous data set and detected in the Gaia-like one are variables with input amplitude just below the amplitude limit. The addition of noise in the light-curve slightly increases their variogram values, and pushes them above the detection limit.
As a conclusion, the variogram method applied to Gaia-like light-curves allows a good recovery of short period variables, with respect to what is expected from the ideal case.

\begin{figure*}[ht]
\centering
\includegraphics[width=0.5\linewidth, page=4, trim={0 0 0 3cm}, clip=true]{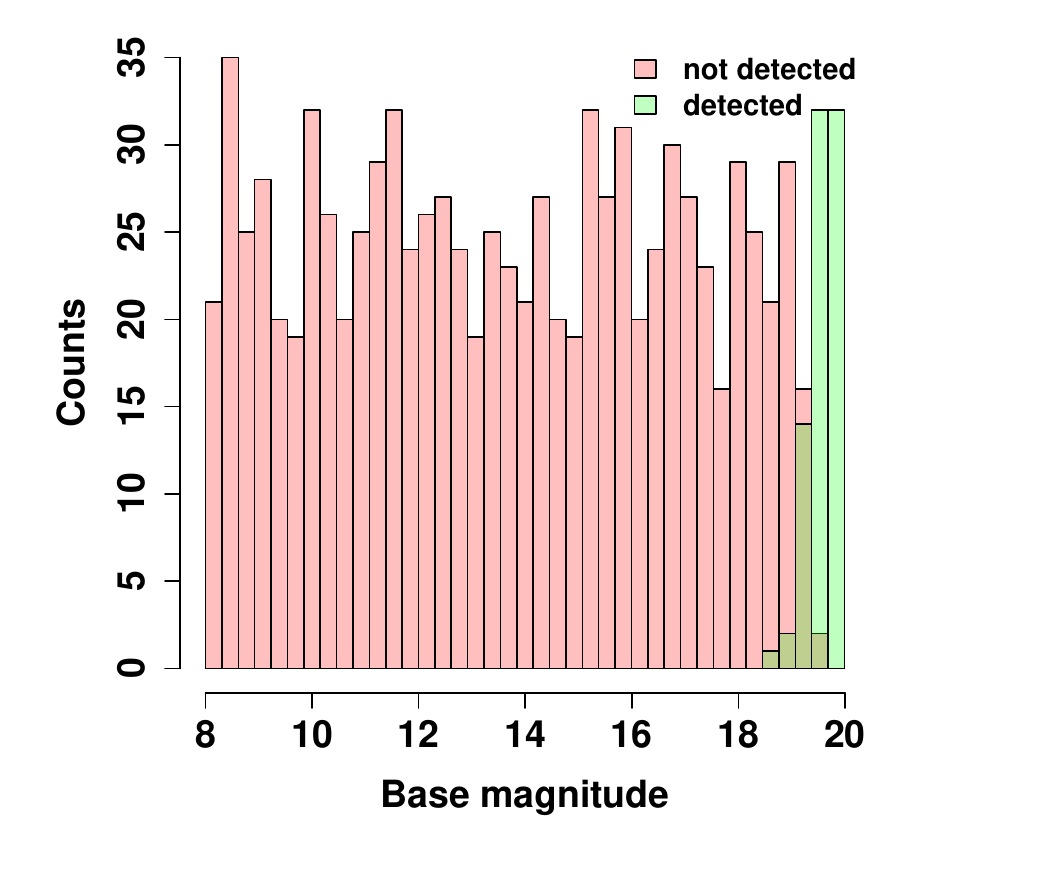}
\caption{Distribution of magnitude for the constant stars simulations, separating the detected (green) and the not detected (red) sources, Gaia-like data set}
\label{fig:histMagConstants}
\end{figure*}

However, ensuring that short timescale periodic variables can be detected by Gaia is necessary, but not sufficient. We also have to make sure that our method limits the number of false detections, for true variables not mixed with an overflow of false positives. Thus, we completed our Gaia-like data set with 1000 simulations of constant stars, with magnitude between $8$ and $20$ mag. Once again, we calculated the associated observational variograms, and applied the detection criterion. Finally, 81 of the simulated constant sources are flagged as short timescale variables, which represents a rate of 8.1\% false positives. As one can see in Figure \ref{fig:histMagConstants}, all the false detections correspond to sources fainter than $\sim 18.5$ mag. At this faint end of Gaia G photometry, the noise measurement level gets close to the limit fixed by our detection threshold. Thus the value of $\gamma_{det}$ used for our analysis is not adapted for faint sources. Hence, the next step will be to refine our choice of $\gamma_{det}$, eventually with different values depending on the average magnitude of the source. The goal will be to find a proper balance between an acceptable rate of false positives, and an optimized detection of as many short period variables as possible.

\section{ePipe: a pipeline for ground-based follow-up of variable stars}
\label{ePipe}

In section \ref{shortTimescaleGaia}, we showed that Gaia data are promising in the exploration of the domain of short timescale periodic variability. By the end of the nominal mission, Gaia will provide several tens of thousands of short timescale variable candidates, together with a lot of homogeneous information about each of these objects, thanks to its simultaneous astrometric, photometric and spectroscopic measurements. However, although Gaia will be very complete, particularly in terms of fraction of variable objects detected, supplementary observations of such candidates will be necessary. It will allow us to confirm the suspected variability and / or to further characterize these variable sources.

In this perspective, we plan to perform photometric ground-based follow-up of short timescale variable stars detected in stellar clusters by Gaia. We will observe them with the Euler telescope in La Silla (Chile), and the Mercator telescope in La Palma (Canary Islands, Spain), ensuring a coverage of both northern and southern skies. During the last few months, we have been developing a pipeline, named \textit{ePipe}, for reduction and analysis of high cadence photometry from Euler and Mercator telescopes. \textit{ePipe} performs classical photometric reduction (i.e. bias substraction and flat field division), as well as astrometry and source detection on the reduced science images. It can also produce aperture photometry for the detected sources, and differential photometry of stellar clusters if a list of reference sources is provided \citep{Saesen2010}.
\textit{ePipe} can process hundreds of observations of stellar clusters or individual sources, and return the corresponding light-curve in apparent magnitude or in differential magnitude. In the future, we will enrich the pipeline with specific tools dedicated to variability analysis: the variogram method, adapted to the specificities of ground-based high cadence follow-up, as well as Fourier techniques for period and amplitude determination.

\begin{figure*}[ht]
\centering
\includegraphics[width=0.7\linewidth]{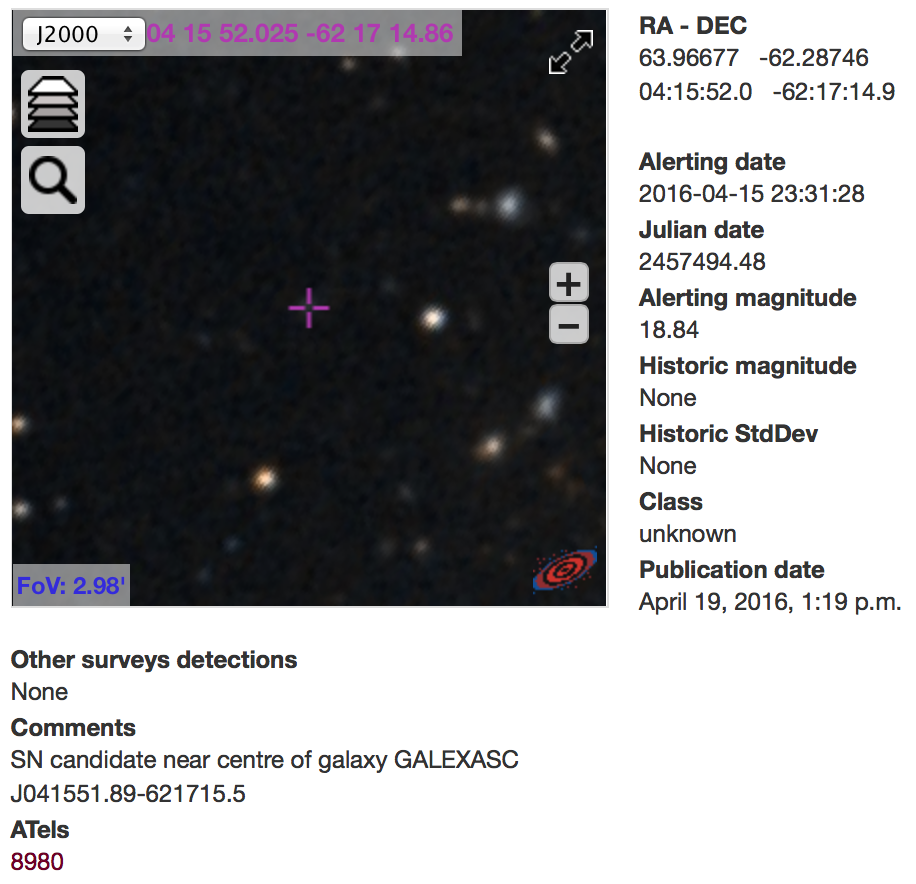}
\caption{SDSS image of the sky surrounding the position of \textit{Gaia16ali}. The position of the alert is indicated by a purple cross. Additional information about the alert is also provided.}
\label{fig:Gaia16aliSDSS}
\end{figure*}

\begin{figure*}[ht]
\centering
\includegraphics[width=0.5\linewidth]{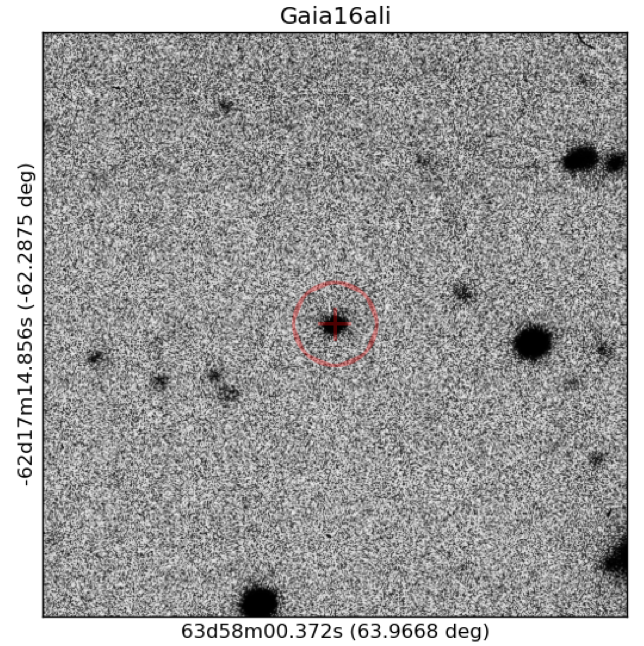}
\caption{ECAM image of the sky surrounding the position of \textit{Gaia16ali}. The position of the alert is indicated by a red circled cross.}
\label{fig:Gaia16aliECAM}
\end{figure*}

\begin{figure*}[ht]
\centering
\includegraphics[width=0.99\linewidth, page=4, trim={1cm 4cm 0cm 4cm}, clip=true]{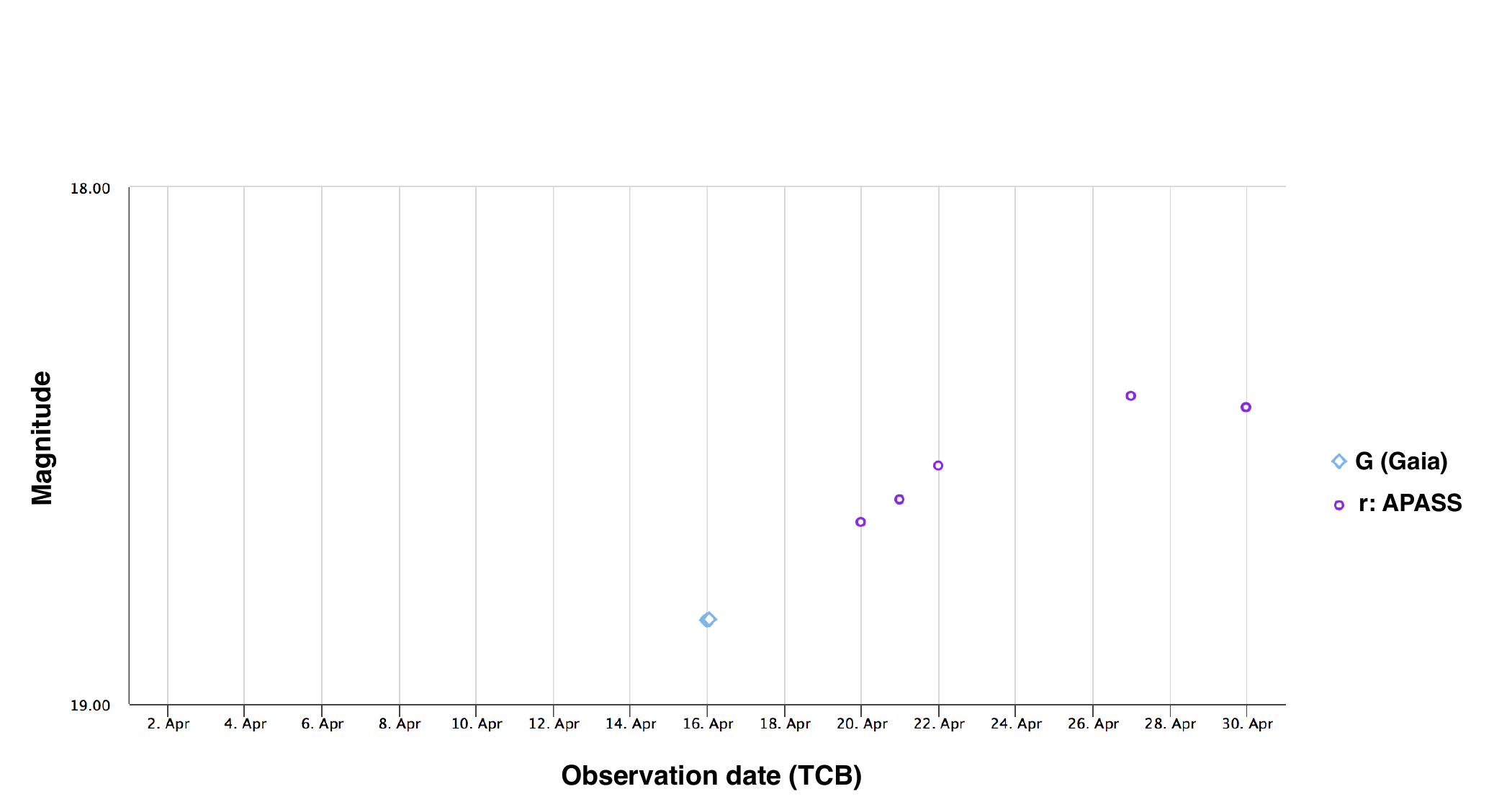}
\caption{Light-curve of \textit{Gaia16ali}, combining Gaia data (in blue) and our reduced ECAM data (in purple).}
\label{fig:Gaia16aliLC}
\end{figure*}

Though we have not found Gaia short timescale variable candidates yet, we already use \textit{ePipe} for the Gaia data, in the frame of the Gaia Science Alerts \citep{Hodgkin2013,Wyrzykowski2016}. As part of Gaia DPAC, the Gaia Science Alerts (GSA) group, mostly based in Cambridge (UK), is in charge of rapid analysis of daily Gaia data deliveries, in order to announce transient objects observed by Gaia. The GSA team identify and classify either new sources which were not visible in previous scans, or old sources experiencing a sudden brightening or darkening. GSA has started routine operation in July 2014, and alerts are publicly reported on the GSA webpage\footnote{\url{http://gaia.ac.uk/selected-gaia-science-alerts}}. To fully exploit their scientify potential, these transient objects require immediate additional observations from the ground, including multiband photometric follow-up. Hence, when an observer involved in the GSA working group has time on a telescope from which a recent alert is visible, the target can be confirmed and characterized doing photometry and spectroscopy. Then, follow-up data are sent for automated calibration to the Cambridge Photometric Calibration Server\footnote{\url{http://gsaweb.ast.cam.ac.uk/followup/}}, and published in the photometric alerts webpage\footnote{hhtp://gsaweb.ast.cam.ac.uk/alerts/home} together with the Gaia data.
In this context, we had several opportunities to observe Gaia Science Alerts during our own observing runs at the Euler and Mercator telescopes. In April 2016, we followed the candidate supernova \textit{Gaia16ali} with the ECAM camera at the Euler telescope in La Silla (Chile). We obtained one image per night with the modified Gunn R passband, during five nights. We reduced each image with \textit{ePipe} immediately after being observed. Figure \ref{fig:Gaia16aliSDSS} shows the SDSS image of the sky around \textit{Gaia16ali}. Figure \ref{fig:Gaia16aliECAM} represents one of our ECAM image of \textit{Gaia16ali} after reduction. Whereas nothing appears at the position of the detected science alert in the SDSS image, a bright source is clearly visible in our ECAM image. Figure \ref{fig:Gaia16aliLC} shows the light-curve of \textit{Gaia16ali} as one can find it in the alert webpage\footnote{\url{http://gsaweb.ast.cam.ac.uk/alerts/alert/Gaia16ali/}}, combining Gaia measurements and our data. Our reaction time was fast enough to the publication of this alert to catch the rising phase of the supernova. With our observations, we confirmed \textit{Gaia16ali} as a supernova candidate, which has been reported in an astronomical telegram \citep[ATel,][]{Roelens2016}\footnote{\url{www.astronomerstelegram.org/?read=8980}}.

\section{Conclusion}
\label{conclu}

In this work, we present our preliminary results for the study of short timescale variability in stellar clusters, combining Gaia data and ground-based follow-ups.

By means of extensive light-curve simulations, we assessed the power of Gaia for detecting variability at typical timescales shorter than a dozen of hours. We showed that, in that prospect, the variogram method applied to the Gaia per-CCD photometry is a promising approach, and ensures the recovery of short period variables with amplitudes above $0.046$ mag ($0.14$ mag for AM CVn stars), with a 8\% rate of false positives. In the future, we will focus on the refinement of our technique to reduce this rate. We already tested the variogram method on some real Gaia light-curves, observed during the first month of nominal science operations, in summer 2014, and more data are currently analyzed.

Once the Gaia short timescale variable candidates will be identified, we will follow-up from ground the members of stellar clusters, firstly to confirm them and validate the future published data in the frame of the DPAC activities, secondly in the perspective of scientific exploitation of Gaia, to further characterize these specific objects. The photometric reduction pipeline we developed in this context, named \textit{ePipe}, is now ready for such a study. We already use \textit{ePipe} in the frame of the GSA photometric follow-up, demonstrating the important synergy between Gaia and ground-based observations. In the future, \textit{ePipe} will implement additional methods specifically dedicated to variability analysis.


\bibliographystyle{cs19proc}
\bibliography{coolStars19_mybib.bib}

\begin{thebibliography}{10}
\providecommand{\natexlab}[1]{#1}

\bibitem[\protect\astroncite{{de Bruijne}}{2012}]{deBruijne2012}
{de Bruijne}, J.~H.~J. 2012, \apss, 341, 31.

\bibitem[\protect\astroncite{{Eyer} \emph{et~al.}}{2015}]{Eyer2015}
{Eyer}, L., {Evans}, D.~W., {Mowlavi}, N., {Lanzafame}, A., {Cuypers}, J.,
  \emph{et~al.} 2015, ArXiv e-prints.

\bibitem[\protect\astroncite{{Eyer} \& {Genton}}{1999}]{Eyer1999}
{Eyer}, L. \& {Genton}, M.~G. 1999, \aaps, 136, 421.

\bibitem[\protect\astroncite{{Hodgkin} \emph{et~al.}}{2013}]{Hodgkin2013}
{Hodgkin}, S.~T., {Wyrzykowski}, L., {Blagorodnova}, N., \& {Koposov}, S. 2013,
  Philosophical Transactions of the Royal Society of London Series A, 371,
  20120239.

\bibitem[\protect\astroncite{{Hughes} \emph{et~al.}}{1992}]{Hughes1992}
{Hughes}, P.~A., {Aller}, H.~D., \& {Aller}, M.~F. 1992, \apj, 396, 469.

\bibitem[\protect\astroncite{{Mignard} \emph{et~al.}}{2008}]{Mignard2008}
{Mignard}, F., {Bailer-Jones}, C., {Bastian}, U., {Drimmel}, R., {Eyer}, L.,
  \emph{et~al.} 2008, In \emph{A Giant Step: from Milli- to Micro-arcsecond
  Astrometry}, edited by W.~J. {Jin}, I.~{Platais}, \& M.~A.~C. {Perryman},
  \emph{IAU Symposium}, vol. 248, pp. 224--230.

\bibitem[\protect\astroncite{{Pojmanski}}{2002}]{Pojmanski2002}
{Pojmanski}, G. 2002, \actaa, 52, 397.

\bibitem[\protect\astroncite{{Roelens} \emph{et~al.}}{2016}]{Roelens2016}
{Roelens}, M., {Blanco-Cuaresma}, S., {Semaan}, T., {Palaversa}, L., {Mowlavi},
  N., \emph{et~al.} 2016, The Astronomer's Telegram, 8980.

\bibitem[\protect\astroncite{{Saesen} \emph{et~al.}}{2010}]{Saesen2010}
{Saesen}, S., {Carrier}, F., {Pigulski}, A., {Aerts}, C., {Handler}, G.,
  \emph{et~al.} 2010, \aap, 515, A16.

\bibitem[\protect\astroncite{{Wyrzykowski}}{2016}]{Wyrzykowski2016}
{Wyrzykowski}, {\L}. 2016, ArXiv e-prints.

\end{thebibliography}

\newpage

\begin{table*}[t]
\centering
\caption{List of periodic short timescale variable types that were simulated}
\label{tab:varTypeListPeriodic}
\begin{tabular*}{\linewidth}{l c l c l c | c |}
\noalign{\smallskip}\hline\noalign{\smallskip}
\bf{Variable type} & \bf{Period range} & \bf{Amplitude range [mag]} & \bf{Description}\\
\noalign{\smallskip}\hline\noalign{\smallskip}
ZZ Ceti & 0.5 - 25 min & < 0.3 & Pulsating white dwarf\\
AMCVn & 5 - 65 min & < 2 & Eclipsing double (semi) degenerate system\\
$\delta$ Scuti & 28 - 480 min & < 0.9 & Pulsating main sequence star\\
$\beta$ Cephei & 96 - 480 min & < 0.1 & Pulsating main sequence star\\
RRab & 0.2 - 0.5 d & 0.2 - 2 & Pulsating horizontal branch star\\
RRc & 0.1 - 0.5 d & 0.2 - 2 & Pulsating horizontal branch star\\
Algol-like eclipsing binary & 0.15 - 0.5d & 0.2 - 1 & Eclipsing binary of type EA\\
Contact eclipsing binary & 0.1 - 0.5 d & 0.15 - 0.5 & Eclipsing binary of type EB or EW\\
\noalign{\smallskip}\hline
\end{tabular*}
\end{table*}

\begin{table*}[t]
\centering
\caption{Comparison of the detection results between the continuous and the Gaia-like data sets.}
\label{tab:continuousVSgaialike}
\includegraphics[width=0.8\linewidth, page=1, trim={5cm 8cm 8cm 8cm}, clip=true]{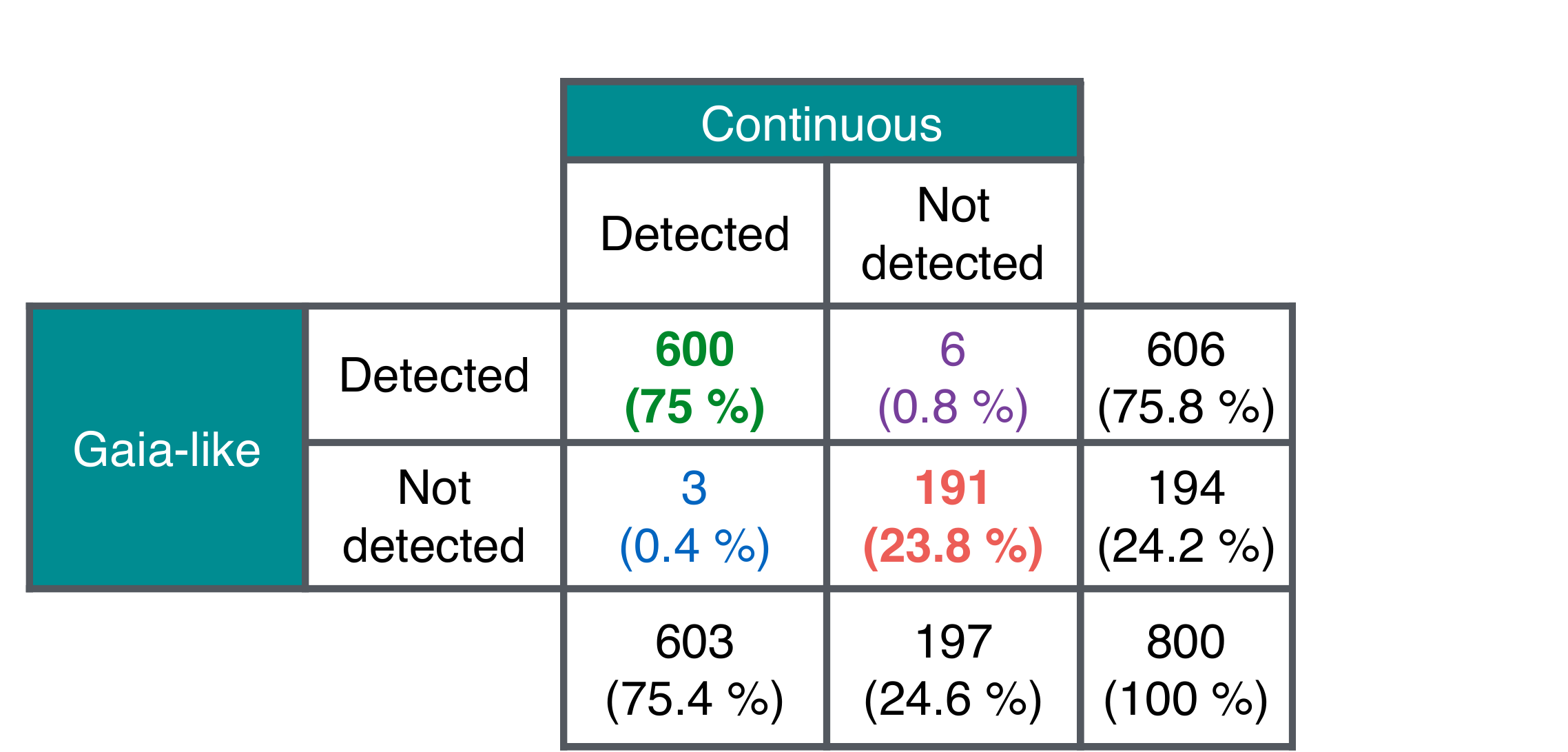}
\end{table*}

\newpage

\end{document}